\documentstyle[12pt,amssymbols]{article}
\title{Classical $R$-matrix structure for the Calogero model}
\author{J.Avan \and M.Talon~~ \thanks{L.P.T.H.E. Universit\'e Paris VI
(CNRS URA 280), Bo\^\i te 126, Tour 16,
$1^{\rm er}$ \'etage, 4 place Jussieu, F-75252 PARIS CEDEX 05} }
\date{October 1992}
\def\x {\stackrel {\textstyle \otimes}{,}}

\begin{document}

\begin{titlepage}
\renewcommand{\thepage}{}
\maketitle
\vskip 2cm

\begin{abstract}
A classical $R$-matrix structure is described for the Lax
representation of the integrable $n$-particle chains of
Calogero-Olshanetski-Perelo\-mov. This $R$-matrix is dynamical, non
antisymmetric and non-invertible. It immediately triggers the
integrability of the Type I, II and III potentials, and the algebraic
structures associated with the Type V potential.
\end{abstract}
\vfill
PAR LPTHE 92/40\hfill Work sponsored by CNRS, France.

\end{titlepage}
\renewcommand{\thepage}{\arabic{page}}

\section{Introduction}
\indent
A general description of classically integrable systems of $n$
particles with two--body interactions $v(x_i - x_j)$ was undertaken in
\cite{OP}. The authors described a set of five two-body potentials,
respectively denoted I $( v(q) = q^{-2})$, II $( v(q) = a^2 {\rm
sinh}^{-2}aq)$, III $(v(q) = a^2 {\rm sin}^{-2} aq)$, IV ($ v(q) = a
{\cal P}(aq, \omega_1 , \omega_2 )$; $ \cal { P}$ being the
double-periodic Weierstrass function), V $( v(q) = q^{-2} + g q^2 )$.
To this list can be added the Toda potential $v(q) = e^{-q}$, denoted
VI. It must be noticed that I, II, III are particular limits of IV.
These potentials were originally derived in \cite{Ca}.

To each simple Lie algebra and choice of one of these six potentials
one can associate a classically integrable system (see Section 2).
The simplest case is the
$A_n$ algebra, for which the first non-trivial Hamiltonian of the
hierarchy describes the original Calogero model:
\begin{equation}
{\cal H} = \sum_{i=1}^{n} p_i^2 + \sum_{i \neq j} v(q_i - q_j)
\label {1.1}
\end{equation}
$(p_i , q_i)$ being canonically conjugate variables for the $i$-th
particle. The choice of Potential V leads to a model which was
recently shown to be related to the collective theory of
$2$-dimensional strings \cite{AJ}.

This Hamiltonian system, and the others obtained from different
choices of algebra (detailed in \cite{OP}), were shown to be
integrable. Their dynamics is described by a Lax representation
\cite{La}, and one is able to construct a set of Poisson-commuting
Hamiltonians from this representation \cite{P4,Wo}. However the
associated algebraic structure of the Lax representation is not known;
in particular the $R$-matrix describing the Poisson	brackets of
the Lax operator \cite{StS}, known to exist from general arguments
\cite{BV}, has not yet been constructed.

Our main result is an explicit construction of the $R$-matrix for the
$A_n$ models with potentials of the type I, II, III and V. The Lax
operator reads:
\begin{equation}
L \equiv \sum_{i = 1}^{n} p_i e_{ii} + i \sum_{ i \neq j} w(q_i -q_j)
e_{ij} \; ; \; v(q_i - q_j) \equiv w^2 (q_i - q_j)
\label{1.2}
\end{equation}
and its $R$-matrix reads (with $(e_{ij})_{kl}=\delta_{ik} \delta_{jl}$):
\begin{eqnarray}
\lefteqn{
R \equiv \sum R^{ijkl} e_{ij} \otimes e_{kl} =}\nonumber\\
&& - \sum_{i \neq j}
\frac{w'(q_i - q_j)}{w(q_i - q_j)} \,\, e_{ij} \otimes e_{ji}
+ \sum_{i \neq j} w(q_i - q_j) \,\, e_{ii} \otimes
(e_{ij} - e_{ji})
\label{1.3}
\end{eqnarray}
Note that $\cal{H}$ in (\ref{1.1}) is ${\rm Tr~} L^2$, an
ad--invariant function of the Lax matrix, hence is conserved.

The plan of this paper is as follows. In Section 2 we recall the main
features of the Lax representation for the
Calogero-Olshanetski-Perelomov models, and the current proofs of
integrability. In Section 3 we derive directly the $R$-matrix
(\ref{1.3}) for the Lax matrix (\ref{1.2}) and check that this leads
to the correct $M$ matrix \cite{OP} in the Lax pair for the
Hamiltonian ${\rm Tr~} L^2$. In Section 4 we elaborate on the case of
Potential V. Two ``Lax matrices'' $L^+$ and $L^-$ are introduced and
the $R$-matrix structure allows an immediate interpretation of the
angle--type variables ${\rm Tr~}( L^{\pm}) ^n$ and action--type
variables ${\rm Tr~} (L^+ L^- )^n$, and of the interesting algebraic
structures arising in the explicit solution of this model, described
in \cite{Pe,AJ2}. Finally we give some concluding remarks on the
extension to the generic potential IV, and other algebras, of this
$R$-matrix formalism.

\section{The Lax representation and integrability of the C. O. P.
model (Type I-IV)}

We shall briefly reproduce here results for the algebra $A_n$
described in detail in the review article
\cite{OP}. Introducing an Ansatz for the Lax pair $(L, M)$ of the form:
\begin{eqnarray}
L_{jk} &=& p_j \delta_{jk} + i(1 - \delta_{jk}) w(q_j - q_k) \nonumber
\\ M_{jk} &=& \delta_{jk} \sum_{i \neq j} z(q_i - q_j) - (1 -
\delta_{jk}) y(q_i - q_k) \nonumber \\ k, j &=& 1 ... n
\label {2.1}
\end{eqnarray}
where $x$ is an odd function and $y, z$ are even functions, it is
shown that the Lax equation $ dL / dt  = [ L , M]$ reproduces
the Hamiltonian equations for the two-body potential $v(q_i - q_j)$,
provided one identifies:
\begin{eqnarray}
v(q) &=& w^2 (q)  \nonumber \\ z(q) = \frac{w''(q)}{2w(q)} &;&
y(q) = -w'(q)
\label{2.2}
\end{eqnarray}
and $w$ obeys the functional equation:
\begin{equation}
w(q_1)w'(q_2) - w(q_2)w'(q_1) = w(q_1 + q_2) (z(q_2) - z(q_1))
\label{2.3}
\end{equation}
This functional equation admits four solutions \cite{Ca,OP2}, one
generic $( w(q) = a\, {\rm sn}^{-1} (aq))$ corresponding to the Type
IV potential, and three limits of this solution, obtained when one of
the periods of sn is sent to $+i \infty$ $(w(q) = a \, {\rm sin}^{-1}
(aq),$ Type III$ ; w(q) = a \, {\rm sinh}^{-1} (aq),$ Type II $)$ and
finally $a$ sent to $0$ $(w(q) = 1/q$, Type I $)$.

The Type V potential $v(q) = 1/q^2 + q^2 $ can be obtained from a
modification of the Lax representation which will be discussed in
Section 4. The Type VI potential $v(q) = e^{-q}$ leads to the open
Toda chain.  Although not belonging to the set defined by the Lax pair
(\ref{2.1}), it nevertheless has some related features which we shall
comment upon in the conclusion.

The Lax representation (\ref{2.1}) is associated with the algebra
$A_n$. A general formulation was discussed in \cite{OP2}, involving
the introduction of general root systems $\Phi$. The two-body potential
becomes:
\begin{equation}
U(\vec{q}) = \sum_{\alpha \in \Phi } g_{\alpha}^2 \, v(q_{\alpha})\; ; \;
q_{\alpha} = (\alpha , \vec{q}\, ) \; ; \;
\vec{q} \equiv (q_1, ... q_n)
\label{2.4}
\end{equation}
The candidate Lax operator reads:
\begin{equation}
L = \sum_{\alpha \in \Phi} g_{\alpha} w(q_{\alpha}) e_{\alpha} -
\sum_{\alpha \in \Phi~{\rm simple}} p_{\alpha} h_{\alpha}
\label{2.5}
\end{equation}
We shall solely describe the details of the derivation for the algebra
$A_n$ and say a few words in conclusion concerning a general algebraic
setting.

Integrability a la Liouville does not automatically follows from the
existence of the Lax representation.  One needs to check that the
conserved quantities ${\rm Tr}_E~ L^n$ have vanishing Poisson brackets
\cite{Ar}.  When an $R$-matrix structure is known, this result is
trivial \cite{Skl1}.
The proof runs as follows:
\begin{eqnarray*}
\lefteqn{
\{\mbox{Tr}_E \, ( L^n ), \mbox{Tr}_E \, ( L^m) \} = \mbox{Tr}_{E \otimes
E} \, \{ L^n  \x L^m\} }\nonumber \\ &=& nm ~\mbox{Tr}_{E
\otimes E} \{L^n \otimes L^m([R , L
\otimes {\bf 1} ] - [R^\Pi ,{\bf 1} \otimes L ])\}
\nonumber \\
&=& 0 \mbox{ by cyclicity of the trace. }
\end{eqnarray*}
However, without such an explicit structure, the
previously proposed demonstrations relied on arguments of asymptotic
behaviour \cite{Mo}, or computational devices using inverse
scattering-type methods \cite{OP2}, or direct --and complicated--
recursion procedures \cite{Wo}.

On the other hand, the Poisson commutation of the ad--invariant
conserved quantities, whichever way it is proved, implies the
existence of an $R$-matrix, a priori dynamical \cite{BV}, which we are
now going to construct.

\section{The $R$-matrix for type I-III}

The Poisson bracket structure for the Lax operator $L$ appearing in
eq.~(\ref{2.1}) is the following:
\begin{equation}
\{ L \x L \} =  \sum_{i \neq j} \frac{w'(q_i -q_j)}{w(q_i- q_j)}
(e_{ii} \otimes e_{ij} - e_{ij} \otimes e_{ii})
\label{3.1}
\end{equation}
where $e_{ij}$ are the generators of $A_n$ in the fundamental
representation.

A generic $R$-matrix structure reads \cite{Skl1,StS,BV}:
\begin{equation}
\{ L \x L \} = [R, L \otimes {\bf 1}] - [R^{\Pi} , {\bf 1} \otimes L]
\label{3.2}
\end{equation}
$R$ belongs to the tensor algebra $A_n \otimes A_n$; $R^{\Pi}$ is the
operator obtained by exchanging the two terms in $A_n \otimes A_n$.
Introducing the components:
\begin{equation}
R \equiv \sum_{a,b,c,d} R^{abcd} e_{ab} \otimes e_{cd}
\label{3.3}
\end{equation}
the Poisson bracket structure (\ref {3.1} -- \ref{3.2}) is equivalent
to the system of overdetermined equations:
\refstepcounter{equation}
\label{3.4}
$$\displaylines{
(p_i - p_a) R^{aicd} + (p_c - p_d)R^{cdai} + \sum_{k} \left(
\frac{R^{akcd}}{q_k - q_i} -\frac{R^{kicd}} {q_a - q_k}
-\frac{R^{ckai}}{q_k - q_d} + \frac{R^{kdai}}{q_c - q_k} \right)
\cr =~ \delta_{ai} (\delta_{ac} - \delta_{ad})
\frac{w'(q_c -q_d)}{w(q_c - q_d)} - \delta_{cd}(\delta_{ac} -
\delta_{ic}) { {w'(q_a - q_i)} \over {w(q_a - q_i)}} \hfill
 (\theequation) \cr  }$$

In order to solve this system, we shall make a number of assumptions
which will ultimately restrict us to considering only the potentials
of type I to III.

\vskip 0.5cm
\underline{Assumption 1} : $R$ does not depend on $p_i$
\vskip 0.5cm

This decouples (\ref{3.4}) into two sets of equations. The first one
is easily solved:
\begin{equation}
(p_i - p_a) R^{aicd} + (p_c - p_d)R^{cdai} = 0
\label{3.5}
\end{equation}
One gets:
\begin{eqnarray}
R^{aicd} = 0 \,\, \forall a \neq i \neq c \neq d &;& R^{cdaa} =
R^{acad} = R^{acda} = 0 \,\,
\forall a \neq c \neq d \nonumber \\
R^{aiia} = -R^{iaai} \,\, \forall a \neq i &;& R^{aiii} = R^{iaii} = 0
\,\, \forall a \neq i
\label{3.6}
\end{eqnarray}

This leaves as free parameters of $R$: $R^{aacd} ; R^{aaii} ,
R^{aiai} , R^{aiia} ; R^{iiai} , R^{iiia} ; R^{iiii}$.  Plugging these
residual terms into (\ref{3.4}) leads to a new set of equations:
\refstepcounter{equation}
\label{3.7}
$$\displaylines{
(a)~ (R^{aacd} - R^{iicd})w(q_a - q_i) = (R^{ccai} - R^{ddai}) w(q_c
- q_d) \hfill \cr
(b)~ R^{ccii} = R^{ddii} \,\, \forall c,d \hfill\cr
(c)~ (R^{adad} + R^{aiai} ) w(q_d - q_i) + (R^{aaad} -
R^{iiad}) w(q_a - q_i) \hfill\cr
\hfill + (R^{aaai} - R^{ddai}) w(q_d - q_a) = 0\cr
(d)~ (R^{adda} - R^{aiia} ) w(q_d - q_i) + (R^{aada} - R^{iida})
w(q_a - q_i) \hfill\cr
\hfill + (R^{aaai} - R^{ddai}) w(q_d - q_a) = 0 \cr
(e)~ -(R^{dada} + R^{iaia} ) w(q_d - q_i) + (R^{aada} - R^{iida})
w(q_a - q_i) \hfill\cr
\hfill + (R^{aaia} - R^{ddia}) w(q_d - q_a) = 0\cr
(f)~ \hphantom{-~} R^{aaca} - R^{aaac} + R^{ccac} - R^{ccca} = 0 \hfill\cr
(g)~ -R^{iiii} - R^{iaai} + R^{aaii} - R^{aiai} = \frac{w'(q_a -
q_i)}{w(q_a - q_i)} \hfill\cr
(h)~ \hphantom{-~} R^{iiii} - R^{iaai} - R^{aaii}
+ R^{iaia} = \frac{w'(q_a - q_i)}{w(q_a - q_i)} \hfill (\theequation)\cr}$$
This system actually admits a solution with a minimal number of
non-zero parameters, obtained by attempting to trivialize a maximal
number of the above equations. Namely one sets:

\vskip 0.5cm
\underline{Assumption 2}:
\begin{equation}
R^{aacd} = 0 \; ; \; R^{ccii} = 0 \; ; \; R^{iiii} = 0 \; ; \;
R^{aiai} = 0
\label{3.8}
\end{equation}
thereby trivializing $(a,b,f,g,h)$ and leaving two non-vanishing
parameters:
\begin{equation}
R^{iaai} = \frac{w'(q_a - q_i)}{w(q_a - q_i)} \; ; \; R^{aada} =
-R^{aaad} = c.w(q_a - q_d)
\label{3.9}
\end{equation}
provided (from $(d)$) that $g \equiv 1/w$ obeys the functional
equation:
\begin{equation}
g(u)g'(v) + g(v)g'(u) = c. g(u+v)
\label{3.10}
\end{equation}
This functional equation is obeyed by the $w$ (or $g$) functions
corresponding to the potentials of Type I, II and III, but not Type
IV, with the constant $c$ equal to 1; hence our two assumptions,
although restrictive, finally lead us to the $R$-matrix structure for
these three potentials :
\begin{equation}
R = -\sum_{i \neq j} \,\, \frac{w'(q_i - q_j)}{w(q_i - q_j)} \,\,
e_{ij} \otimes e_{ji} + \sum_{i \neq j} w(q_i - q_j) \,\, e_{ii}
\otimes (e_{ij} - e_{ji})
\label{3.11}
\end{equation}
with $w(q) = 1/q $\, (Type I), $\, 1/{\rm sin}(q)$ (Type II),
$1/{\rm sinh}(q)$ (Type III).  This $R$-matrix is dynamical,
non-antisymmetric and non-invertible.

{}From the form of the $R$-matrix (\ref{3.11}) immediately follows the
$M$ matrix of the Lax pair associated to any Hamiltonian of the
hierarchy. In particular for ${\cal H} = {\rm Tr}\, (L)^n$, one has
\cite{StS,BV}:
\begin{equation}
M = {\rm Tr}_2\,  (R.d{\cal H}) = {\rm Tr}_2\, (R.{\bf 1} \otimes L^{n-1})
\label{3.12}
\end{equation}
In the case $n=2$, when ${\cal H}$ is given by (\ref{1.1}), one gets
back the Olshanetski-Perelomov formula \cite{OP}:
\begin{equation}
M = \sum_{i \neq j} \,\, \frac{w'(q_i - q_j)}{w(q_i - q_j)} \,\,
e_{ij} + \sum_{i \neq j} w^2 (q_i - q_j) e_{ii}
\label{3.13}
\end{equation}

Moreover the remarkable identities which allowed to show the Poisson
commutation of the eigenvalues of $L$ are naturally equivalent to the
existence and structure of the $R$-matrix, which is after all a
compact canonical way of formulating special Poisson bracket relations
leading to integrability.

\section{The Type V Potential}

A related Poisson structure can now be described for the Type V
potential $v(q) = q^{-2} + gq^2$. The Lax representation \cite{P3} is
not isospectral; one introduces two Lax operators $L^+$ and $L^-$,
respectively:
\begin{equation}
L^{\pm} = L \pm i \omega  \, {\rm Diag}\,(q_i) \; ; \; L
\equiv L(w_I \equiv 1/q) \; ; \; g \equiv \omega^2
\label{4.1}
\end{equation}

They obey modified Lax equations:
\begin{equation}
\frac{ dL^{\pm}}{dt} = [L^{\pm} , M(w_I)] \pm g L^{\pm}
\label{4.2}
\end{equation}
from which one defines \cite{Pe} angle--type variables ${\rm Tr~}
(L^{\pm})^n$, and action variables ${\rm Tr~} (L^+ L^-)^n $.

The Poisson algebra of these variables is a consequence of the
$R$-matrix structure of $L^+$ and $L^-$. One ends up with:
\begin{eqnarray}
\{ L^+ \x L^+ \} &=& [R, L^+ \otimes {\bf 1}] -[R^{\Pi} ,
{\bf 1} \otimes L^+] \label{4.3} \\
\{ L^- \x L^- \} &=& [R, L^- \otimes {\bf 1}] -[R^{\Pi} ,
{\bf 1} \otimes L^-] \label{4.4} \\
\{ L^+ \x L^- \} &=& [R, L^+ \otimes {\bf 1}] -[R^{\Pi} ,
{\bf 1} \otimes L^-] \nonumber \\
&+& i \omega \,\biggl( \, \sum_{i \neq j} e_{ij} \otimes e_{ji} + \sum_{k}
e_{kk} \otimes e_{kk}\, \biggr) \label{4.5}
\end{eqnarray}
{}From (\ref{4.3}, \ref{4.4}) it immediately follows that the angle
variables $B^{\pm}_n \equiv Tr (L^{\pm})^n$ Poisson--commute when
having the same $\pm$ gradation. From (\ref{4.5}) one also deduces,
not so straightforwardly, that the conserved quantities ${\rm Tr~} (L^+
L^-)^n $ Poisson--commute. Precisely one has:
\refstepcounter{equation}
\label{4.6}
$$\displaylines{
\{ L^+ L^- \x L^+ L^- \} =\hfill\cr
[R.{\bf 1} \otimes L^- + {\bf 1}
\otimes L^+ .R ,  L^+ L^- \otimes {\bf 1}]
- [L^+ \otimes {\bf 1} . R^{\Pi} + R^{\Pi} . L^- \otimes {\bf 1} ,
{\bf 1} \otimes L^+ L^-] \hfill\cr
+ i \omega \,\, {\bf 1} \otimes
L^+ . \Pi_{sl(n)} . L^- \otimes {\bf 1} - i \omega \,\, L^+ \otimes
{\bf 1} . \Pi_{sl(n)} .  {\bf 1} \otimes L^- \hfill (\theequation)
\cr }$$
where $ \Pi_{sl(n)} \equiv \sum_{i \neq j} e_{ij} \otimes e_{ji} +
\sum_{k} e_{kk} \otimes e_{kk}$. Incidentally the $R$-matrix structure
in the first two terms of (\ref{4.6}) is a nice example of a second
Poisson structure canonically obtained from a first structure by a
Sklyanin-type bracket \cite{Skl2}.

Hence in computing the brackets $\{ {\rm Tr}~ (L^+ L^-)^n ,
{\rm Tr}~ (L^+ L^-)^m \}$
the contribution from the commutators vanish, as usual in such
computations, and one is left with:
\refstepcounter{equation}
\label{4.7}
$$\displaylines{
\{ {\rm Tr}~ (L^+ L^-)^n , {\rm Tr}~ (L^+ L^-)^m \} = + i \omega L^-
(L^+ L^-)^{n-1} \otimes (L^+ L^-)^{m-1} L^+ .\Pi_{sl(n)}
\hfill\cr
- i \omega \, (L^+ L^-)^{n-1} L^+ \otimes L^- (L^+ L^-)^{m-1}
.\Pi_{sl(n)} \hfill (\theequation)\cr}$$
Clearly the r.h.s. of (\ref{4.7}) is an odd function of $\omega$ since
$L^+ \leftrightarrow L^- $ when $ \omega
\leftrightarrow - \omega$. However  ${\rm Tr~} (L^+ L^-)^n  =
Tr (L^- L^+)^n$ by cyclicity and therefore the l.h.s. of
(\ref{4.7}) is an even function of $\omega$ . It follows that both
sides must be equal to $0$. Hence the conserved quantities ${\rm Tr~} (L^+
L^-)^n $ Poisson--commute and the potential V is Liouville--integrable.

Some comments are in order at this point. The Type-V potential has
received a lot of attention recently. It is a discretized version of
the collective field theory for two-dimensional strings \cite{AJ}.
Clearly the diagonalization of this field theory, undertaken in
\cite{AJ2}, and the underlying $w_{\infty}$ algebra of
eigen-operators, is a natural consequence of the algebraic structure
(\ref{4.3} -
\ref{4.5}). The operators $B^{\pm}_n$ are in fact ``shifted" action
variables of the potential type I ; they realize through the crossed
terms of (\ref{4.5}) an algebra of oscillator type, which then
generates naturally the $w_{\infty}$ algebra \cite{Bak}. Moreover they
are natural angle-type variables of the Type V potential, hence they
diagonalize the Hamiltonian ${\rm Tr~} (L^+ L^-)$.

Finally a recent resolution of the Calogero model \cite{Br} relies on
the existence of a very similar algebra of covariant derivatives a la
Knizhnik-Zamolodchikov. However an exact relation between these two
algebras is not yet available.

\section{Conclusion}

As we have mentioned, this construction does not realize an $R$-matrix
for generic (Type IV) elliptic potentials, due to the functional
constraint (\ref{3.10}). From the derivation in Section 3, it is clear
that the $R$-matrix must then acquire supplementary terms , for
instance $R^{iikl}$, making an eventual complete resolution rather
complicated. We shall not attempt to discuss it any further here.

The extension to other algebras sounds more promising. As indicated,
the generic Lax operator takes the form:
\begin{equation}
L = \sum_{\alpha \in \Phi} g_{\alpha} w(q_{\alpha}) e_{\alpha} -
\sum_{\alpha \in \Phi  {\rm simple}} p_{\alpha} h_{\alpha}
\label{5.1}
\end{equation}
{}From our previous results, and allowing for general Chevalley
relations $$[e_{\alpha} , e_{\beta}] = N_{\alpha \beta} \,\, e_{\alpha
+ \beta}$$ we expect for the Type I-III potential an $R$-matrix of the
form:
\begin{equation}
R = \sum_{\alpha \in \Phi} c(\alpha) \frac{w'(\alpha (q))}{w(\alpha(q))}
e_{\alpha} \otimes e_{- \alpha} + \sum_{\alpha \in \Phi} d(\alpha)
w(\alpha (q)) h_{\alpha} \otimes e_{\alpha}
\label{5.2}
\end{equation}
Here $c(\alpha) , d(\alpha)$ are constant $(q)$ functions of the
particular root $\alpha$.

Finally it is interesting to note that Type VI potential $(x(q) =
e^q)$ , although having a very different Lax formulation, has an
associated $R$-matrix \cite {Tod} reminding of (\ref{3.11}). Provided
one eliminates the second term, and replaces $x(q)$ by $e^q$ in
(\ref{3.11}), one ends up with:
\begin{equation}
R = \sum_{ i > j} e_{ij} \otimes e_{ji} - \sum_{ i < j} e_{ij} \otimes
e_{ji} \equiv R^{\rm Toda}
\label{5.3}
\end{equation}

{\bf Acknowledgements}

We wish to thank Olivier Babelon, John Harnad, Alexei Reiman, and Claude
Viallet for fruitful discussions.

\end{document}